\documentclass[tradiabstract]{aa} 
\usepackage{graphicx}
\usepackage{natbib}
\usepackage{times}

\begin{document}

\title{Modelling solar irradiance variability on time scales
       from minutes to months}

\author{Andrey D. Seleznyov\inst{1}
       \and Sami K. Solanki\inst{1,}\inst{2}
       \and Natalie A. Krivova\inst{1}}

\offprints{S.\,K. Solanki}

\institute{\inst{} Max-Planck-Institut f{\"u}r Sonnensystemforschung,
           37191, Katlenburg-Lindau, Germany
           \and
           \inst{} School of Space Research, Kyung Hee University,
           Yongin, Gyeonggi 446-701, Korea}

\date{Received October 14, 2008; accepted }

\abstract{We analyze and model total solar irradiance variability on time
scales from minutes to months,
excluding variations due to p-mode oscillations,
using a combination of convective and magnetic components.
These include granulation, the magnetic network, faculae and sunspots.
Analysis of VIRGO data shows that on periods of a day or longer solar
variability
depends on magnetic activity, but is nearly independent at shorter periods.
We assume that only granulation affects the solar irradiance variability on
time scales from minutes to hours.
Granulation is described as a large sample of bright cells and dark lanes
that evolve according to rules deduced from observations and radiation
hydrodynamic simulations.
Comparison of this model
combined with a high time resolution magnetic-field based irradiance
reconstruction,
with solar data reveals a good correspondence except at periods of 10 to 30
hours.
This suggests that the model is missing some power at these periods, which
may be due to the absence of supergranulation or insufficient sensitivity of
MDI magnetograms used for the reconstruction of the magnetic field-based
irradiance reconstructions.
Our model also shows that even for spatially unresolved data (such as those
available for stars) the Fourier or wavelet transform of time series sampled
at high cadence may allow properties of stellar granulation, in particular
granule lifetimes to be determined.
}


\authorrunning{Seleznyov et al.}
\titlerunning{Modelling of solar irradiance variability}

\maketitle

\section{Introduction}
\label{intro}

Models of solar irradiance on time scales of days to the solar cycle
have reached a certain maturity. They reproduce the observations
with high accuracy \citep[e.g.,][and references
therein]{solanki-et-al-2005a,solanki-krivova-2006,krivova-et-al-2011a}.
Shorter time scales have been dealt with much more
summarily.
Traditionally, interest in time scales of minutes to days
has derived from helioseismology (and more recently
asteroseismology) since the Sun's `noise' background produced by
convection and magnetism is a limiting factor in detecting
oscillation modes \citep{harvey-duvall-84,andersen-et-al-94,rabello-et-al-97}.
In recent years variability on hours to days time
scales has become important in connection with extrasolar planet
transit detection programs. Stellar noise at these time scales is
the factor finally limiting the size of the planets that can be
detected with this technique \citep{carpano-et-al-2003,aigrain-et-al-2004}.

Here we model solar total irradiance on time scales
of minutes to months.
However, we concentrate on what helioseismologists call `solar noise' and
explicitly do not consider irradiance fluctuations caused by p-modes.
Traditionally, the influence of granulation,
mesogranulation and supergranulation is considered together in
highly parameterized models, while the magnetic field is not taken
into account explicitly
\citep{harvey-duvall-84,andersen-et-al-94,rabello-et-al-97,aigrain-et-al-2004}.
Here we take a
different approach, assuming that all solar variability not due to
oscillations is produced by magnetism and granulation. We compute
the variability and compare it with observations, mainly recorded
by VIRGO on SoHO and TIM on SORCE.
This test should then show if the assumption is justified.
If not then it suggests that some convective or magnetic component is
missing.
Such modelling also provides the basis of using stellar
power spectra to infer the convective and magnetic properties of
stars, since it allows us to explore the diagnostic capabilities
of such spectra.

We model the influence of the granulation in an explicit manner,
although using an empirical model. Main granular characteristics
such as size, lifetime, contrast are based on observational data.
Modelled power spectra of the irradiance caused by the granulation
are compared with the corresponding spectra of the VIRGO data.
Not surprisingly, it is found that total irradiance variability on time
scales longer than a few hours is not well reproduced by the granulation
alone.
We therefore also apply a model of solar
irradiance variations based on the evolution of the solar surface
magnetic field, which reproduces solar total and spectral irradiance
changes on time-scales from days to years. This model is extended to
shorter time scales down to an hour (the shortest time scale on
which MDI magnetograms are available for an uninterrupted interval
of multiple months). Finally we combine irradiance variations caused
by solar magnetic field changes with those caused by granulation.

\section{Sources of irradiance variations}
\label{observed}

We first analyze the observed irradiance variations on time scales
of minutes to a month. For this we mainly employ data obtained by
the Variability of Irradiance and Gravity Oscillations instrument
\citep[VIRGO;][]{froehlich-et-al-95,froehlich-et-al-97} on SoHO.
These data have
high relative accuracy and are recorded at a 1-minute cadence. They are
thus best suited for our purpose. We consider the total irradiance
(version tsi\_d\_v4\_90)
as well as the measurements in the 3 VIRGO spectral channels: red,
green and blue centered at 862~nm, 500~nm and 402~nm, respectively 
\citep[version spma\_level2\_d\_2002;][]{froehlich-2003}.
Detrended data are used, in order to remove the large trends
introduced by degradation of the photometers in the color
channels. This mainly affects very low frequencies, which are not
of interest here. 
We also fill in the numerous gaps in the
1-minute sampled data by linearly interpolating across them. Most
gaps are only a few minutes long and the interpolation should
influence the results mainly at the highest frequencies.

We also consider data from the Total Irradiance Monitor
\citep[TIM;][]{kopp-lawrence-2005,kopp-et-al-2005b}
onboard the Solar Radiation and Climate Experiment
\citep[SORCE;][]{woods-et-al-2000b,rottman-2005} satellite.
SORCE data for the year 2003 sampled each 6 hours were taken from the LASP
(Laboratory for Atmospheric and Space Physics, Boulder, Colorado) Data
Product web page: http://lasp.colorado.edu/sorce/tsi\_data.html.
Following the procedure for the VIRGO data, gaps were filled in using linear
interpolation.

   \begin{figure}
   \centering

   \includegraphics[width=9cm]{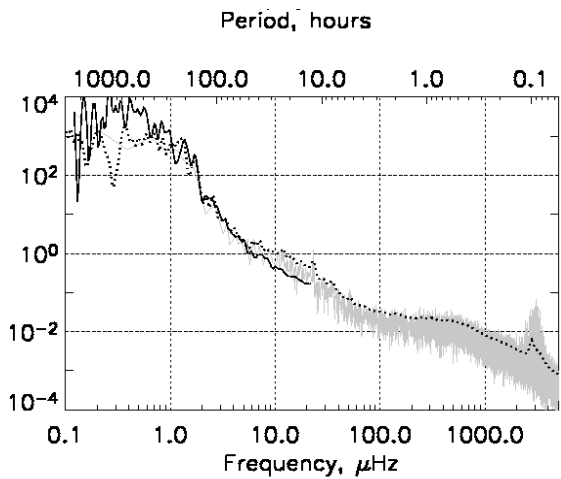}
      \caption{
Fourier (grey line) and global wavelet (black dotted line) power spectra 
(in ppm$^2/60~\mu$Hz)
of the VIRGO data set for the year 2002 sampled at a 1 minute cadence. Black
solid line shows the global wavelet spectrum of the SORCE TIM data for the
year 2003 sampled every 6 hours.
              }
         \label{power_obs}
   \end{figure}

We have applied Fourier and Morlet wavelet transforms to the data.
Both gave essentially the same results (see Fig.~\ref{power_obs}), except
that the global wavelet power spectrum shows smaller fluctuations due to the
smoothing introduced by Morlet wavelets.
The peak in the power at about 5 minutes is due to low degree {\em p}-modes,
eigen oscillations of the Sun.
VIRGO data sampled every minute or every hour (not shown in the figure)
display enhanced power at frequencies above 5~$\mu$Hz relative to SORCE/TIM
data.
This enhancement of power in the VIRGO data is due to 
noise in the electronic calibration.
This noise is roughly constant at a level of
50~ppm at low frequency, but drops at higher frequencies, having its 3-db
point at around 10~$\mu$Hz, which makes it appear as a bump in the solar
spectrum
(C.~Fr{\"o}hlich 2008, priv. comm.).
Differences between VIRGO and TIM data at frequencies below 1~$\mu$Hz are
due to the different times they refer to (2002 for VIRGO and 2003 for TIM).

   \begin{figure}
   \centering
   \includegraphics[width=9cm]{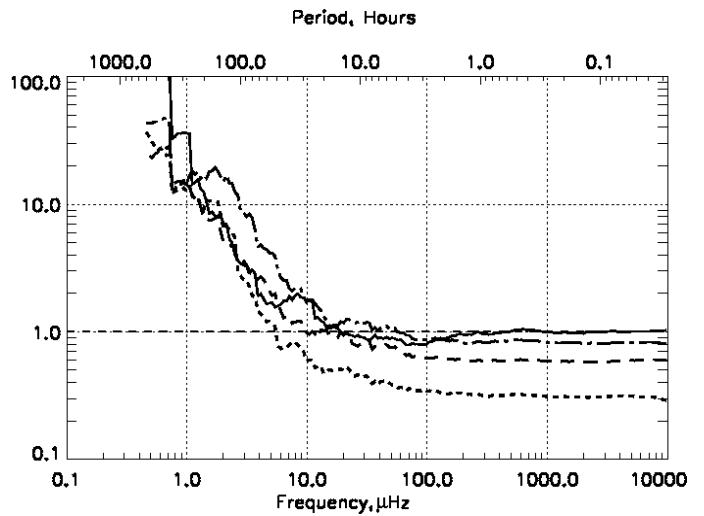}
      \caption{Ratio of wavelet global power spectra of
activity maximum to those at activity minimum for the TSI (solid
line) and three VIRGO channels: red (dot-dashed line), green
(dashed) and blue (dotted).
              }
         \label{wavelet_ratio}
   \end{figure}
%

In order to estimate the time scale at which the relative
contribution of magnetic field evolution (which is solar cycle phase
dependent) and of convection (which is almost independent) becomes
equal, we have analyzed VIRGO records for quiet (1996--1997) and
active (1999--2000) Sun periods individually. For the component
related to active region magnetic fields we expect stronger
variations at activity maximum than at minimum.

The ratio of the wavelet power spectrum of the 1999--2000 period
to that of 1996--1997 is shown in Fig.~\ref{wavelet_ratio}. The
ratio exceeds unity at periods longer than a day. Obviously at
these periods magnetic fields dominate the variability. At shorter
periods, the ratio is essentially 1 in the total irradiance.
Values below 1 are seen in the 3 color channels. They may be an
instrumental artifact, although the reason has not been identified
(C.~Fr{\"o}hlich 2003, priv. comm.).
\citet{froehlich-lean-2004} employed the data between 2000.8 and
2002.8 for the active period to arrive at a value somewhat higher than 1.
Since the Sun was more active during this latter period than during
1999--2000, this may partly explain the higher value.
They have also apparently used a later version of the SPM data.
From this diagram alone it is not possible to say whether there is a
non-magnetic source of the irradiance variations above 10~$\mu$Hz, e.g.
convection, or a magnetic source, e.g. the network, which does not change
strongly in strength over the solar cycle (Harvey 1994).
At frequencies between approximately 10 and 100~$\mu$Hz (which are of
particular interest for planetary transits and g-mode search)
both magnetic field and convection may contribute to the power.

\section{Short term variability: Granulation model}
\label{gran_model}

\subsection{Description of the Model}
\label{model}

To extrapolate from the Sun to other stars it is 
necessary to construct models that are easily scalable to other stars.
Here we concentrate on convection.
Of the main scales of solar convection, there is no evidence that the larger
scales (meso- and supergranulation) show any intrinsic brightness contrast
after the contribution from magnetic fields is eliminated.
Mesogranular structure is best visible when following `corks'
over an hour or two \citep{november-simon-88,november-89}
or when considering the spatial distribution of, e.g.,
exploding granules \citep{roudier-muller-87,ploner-et-al-2000}.
There is no evidence that
they show any intrinsic brightness
contrast
\citep[cf.][]{straus-bonaccini-97,hathaway-et-al-2000,rieutord-et-al-2000,%
roudier-et-al-2009}.
Supergranulation is mainly evident in intensity images 
that sample wavelengths at which small-scale magnetic elements
are bright \citep[e.g.,][]{solanki-93}.
\citet{rast-2003a} has set tight limits on the visible contrast between
supergranule cell interior and boundaries if the magnetic features are
masked out.
If correct, any significant contribution to irradiance variations produced by
a supergranule must come from the evolution of the magnetic field at its
boundary.
We therefore concentrate here on modelling the granulation and deal
separately with the magnetic field (Sect.~\ref{magnetic}).

%
   \begin{figure}
   \centering
   \includegraphics[width=9cm]{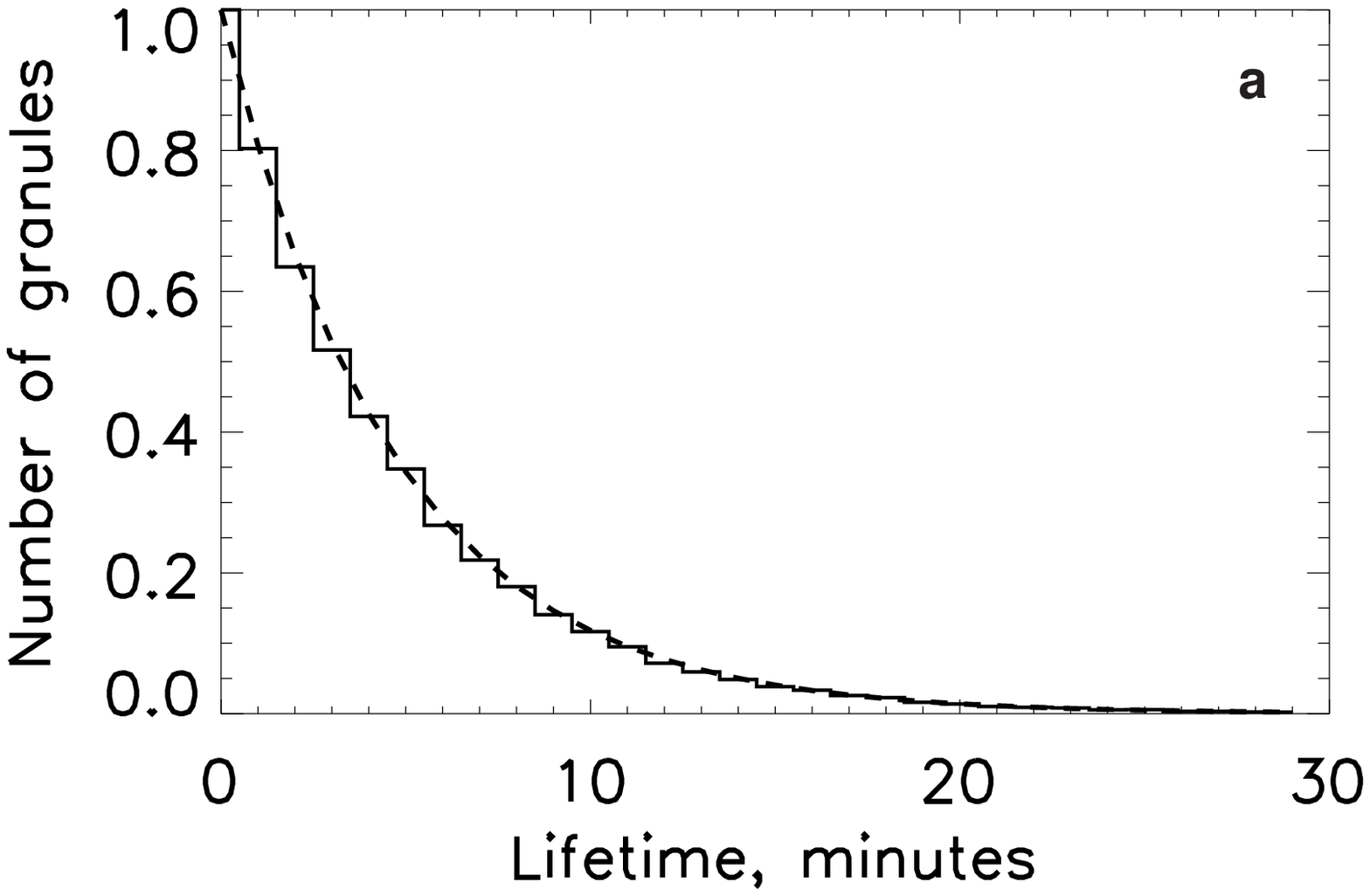}\hspace{5mm}
   \includegraphics[width=9cm]{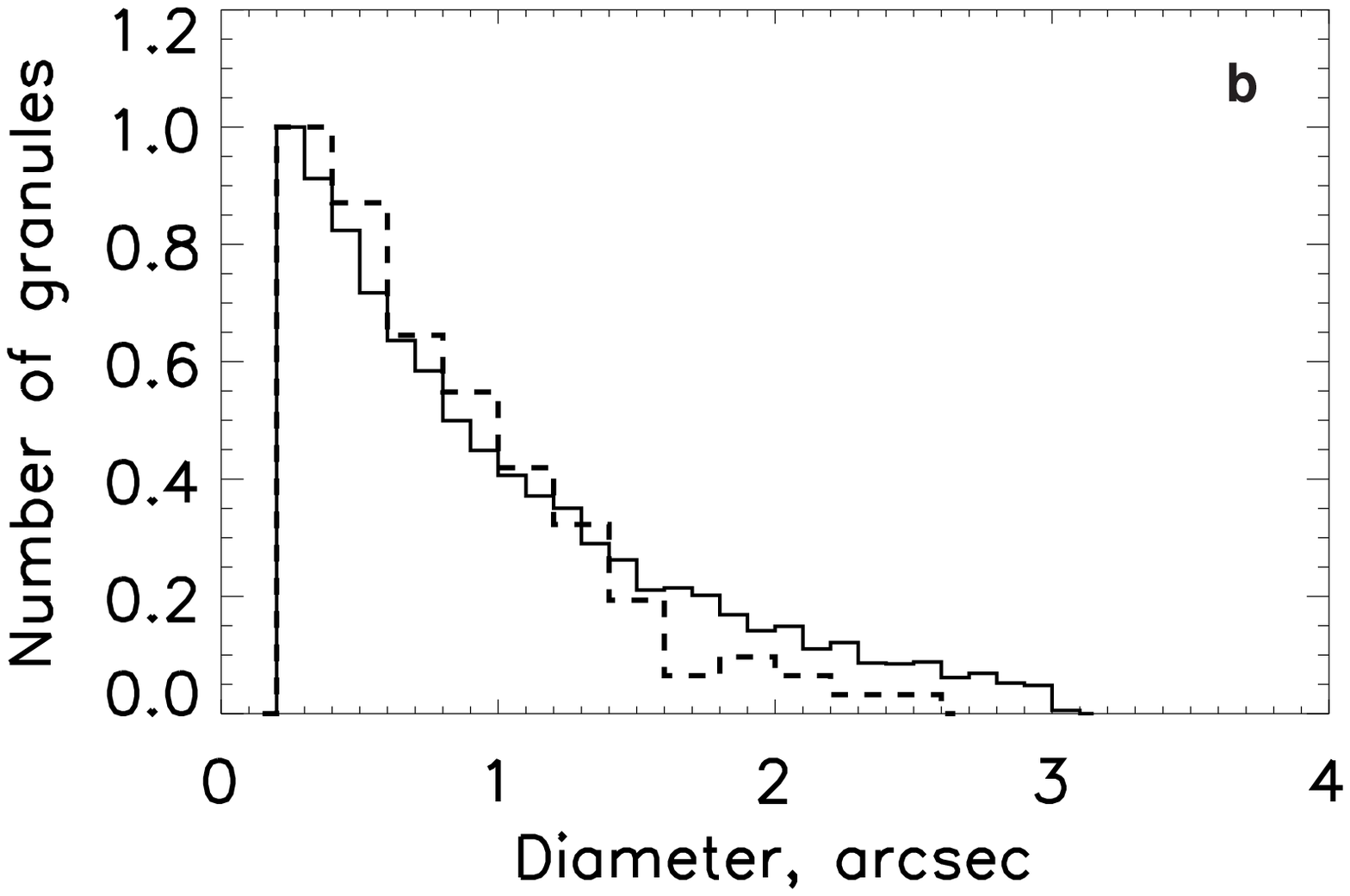}
      \caption{Normalized distributions: a) granule lifetimes;
b) granule sizes. Plotted is a snapshot at a moment of the evolution.
Dashed lines give observed distributions, solid lines the distributions in
the `standard' model. }
         \label{distr}
   \end{figure}
%
%

To model the influence of the roughly $10^6$ granules on
the solar disk, we use a simple
parameterized description of individual granules and a statistical
approach to the evolution of the whole ensemble.
Each granule is treated as a brightness structure, with a
time-dependent bright part (granule) that is assigned a brightness value and
area and a dark part (intergranular lane), whose brightness is fixed and
does not change with time.
The granule distribution and evolution are determined by the following free
parameters:
the granule expansion or contraction rate, the granule lifetime,
rate of brightness increase or decrease, ratio of the splitting granules to
ones that dissolve.

The key features of our model are:
1) The main birth and death mechanisms of granules are
fragmentation (birth and death) and emergence from (birth), or dissolution
into (death) the background
\citep[e.g.,][]{mehltretter-78,kawaguchi-80,hirzberger-et-al-99a}.
2) The brightness value of each granule is assigned randomly
within a given range.
3) The lifetime distribution of all
granules follows an exponential law (Fig.~\ref{distr}a). For the
models designated as the `standard' model the lifetime
distribution follows the one derived empirically by
\citet{hirzberger-et-al-99a} with the decay time of $\tau=4.68$~minutes.
4) The initial distribution of granule sizes in the standard model is
described by the best fit exponential function to
the observed distribution published by \citet{roudier-muller-87}.
The observed distribution is represented by the dashed line in
Fig.~\ref{distr}b.

While granules evolve their size changes such that larger granules
expand and smaller ones shrink.
Due to the excess pressure large granules
build up in the photosphere relative to their neighbors
\citep[e.g.,][]{ploner-et-al-99}.
At the end of its life, depending on its size, each granule can either split
into two parts, thus forming 2 new granules, or dissolve into the
background.
The end of a granule's life is reached either when the
preassigned time (lifetime) has run out or when its size reaches
the limits of the size distribution. The rate at which the
granule's size changes grows linearly from 0 in the middle of the
size distribution to the maximum of 0.1 arcsecond per minute at
the lowest and highest diameter distribution edges. This
effectively means that the smaller the granule the faster it is
squeezed out of existence by its neighbors, and the bigger the
granule the faster it reaches the upper size limit imposed in this
model and splits into two parts. 
The individual
areas of the child granules lie randomly between 1/4 and 3/4 of
the area of the parent granule.

We attempt to reduce sudden jumps
in the brightness at the time of death or birth of a granule
(which produce artificial high-frequency power). Thus the
brightness of granules born out of the splitting of a larger
granule slightly increases, in order to compensate for the
increased area of the intergranular network, which after splitting
surrounds 2 granules instead of one before splitting.
Small granules that are close to disappearing or have just been born have
brightness levels close to the background intergranular lanes.
Such transitions are linear and
limited in time: brightness drops or rises are completed in 3
minutes.
The total number of granules is kept constant by maintaining a
balance between appearing and dying granules.
Also, the size distribution is roughly maintained by allowing fresh granules
with the appropriate size distribution to emerge from the intergranular
lanes.
In Fig.~\ref{distr}b we plot the empirical size distribution of granules
(dashed curve), which is very close to the initial size distribution.
The solid line represents the size distribution of the synthetic granules at
a typical later time.
The output irradiance of the model is normalized by the total area
of the sun at each time step.

In order to check the importance of model parameters, we have
varied each of them individually, while keeping the rest fixed.
The results of this parameter study are presented in Sect.~\ref{results}.
Two other parameters, the total number of granules, $N_{tot}$, and the
average brightness contrast between granules and intergranular lanes move the
irradiance power spectrum up or down, but do not influence its slope. They
are therefore not discussed further here.

%
%
   \begin{figure}
   \centering
 \includegraphics[width=9cm]{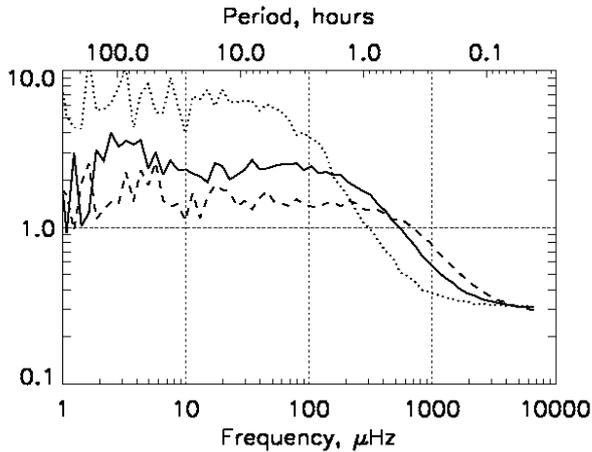}
      \caption{
 Wavelet power spectra 
(in ppm$^2/60~\mu$Hz)
 for different granule lifetimes.
 Solid line shows the result based on the solar lifetime distribution with a
 mean lifetime of $\sim$5.2 min.; dotted line is for longer lifetimes with a
 mean of $\sim$23.7 min. and dashed line for shorter lifetimes with a mean
 of $\sim$2.9 min.
 }
         \label{modelvar1}
   \end{figure}

%
%

\subsection{Results}
\label{results}

\begin{figure}
   \centering
\includegraphics[width=9cm]{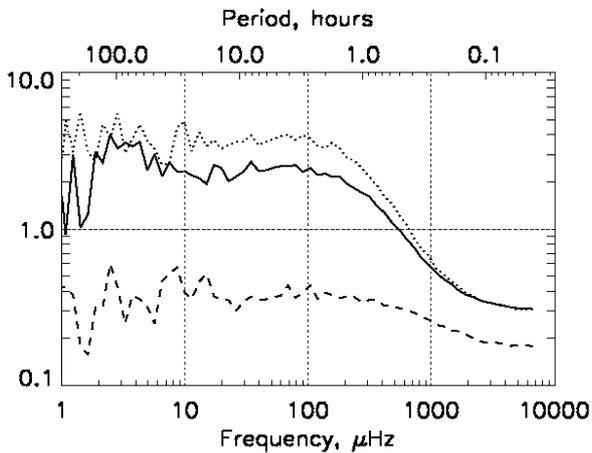}\hspace{5mm}
      \caption{Wavelet power spectra
(in ppm$^2/60~\mu$Hz) for different granule
 sizes (solid line: maximum diameter of a granule is 3 arcsec,
 dotted line: maximum diameter is 9 arcsec, dashed line: maximum diameter 2
 arcsec).
}
         \label{modelvar3}
   \end{figure}

Power spectra from a parameter study carried out with our model are
shown in Figs.~\ref{modelvar1} and \ref{modelvar3}.
Except for the parameter explicitly
mentioned in each case, all others
remain unchanged at their standard values:
mean granule lifetime is approximately 5 minutes, mean
granule diameter is approximately 0.9 arcseconds.

At low frequencies all model
power spectra are relatively independent of frequency (but
with increasing fluctuations towards lower frequency due to the fewer
periods of this length sampled by the simulation).
Above a certain frequency the power drops approximately as a power law.
The flattening of the power spectrum at higher frequencies is due to
aliasing introduced by the fact that granules evolve also on periods shorter
than the employed time step (1 minute); imposed by the typically 6 months
solar time that a model has to be run in order to obtain reliable results
also at lower frequencies.

The curves in Fig.~\ref{modelvar1} refer to different average
granule lifetimes. 
As expected, the frequency at which the power starts to decrease
(i.e. the `knee' in the wavelet power spectrum)
increases with decreasing lifetime
of granules. The corresponding period roughly doubles as the average
granule lifetime is doubled, so that this parameter is potentially a
diagnostic of stellar granule lifetimes. Note, however, that the
period of the `knee' is roughly 10 times longer than the average
granule lifetime.


In Fig.~\ref{modelvar3} the diameter of granules has been varied.
The number of granules is kept constant, so that the total area
covered is larger in the case of larger granules (corresponding to
a bigger star). Obviously small granules produce less noise than
big ones with a shallower slope of the power below several hours.
This is caused by the fact that the ratio of the granule area to
that of the surrounding intergranular lanes is higher for big
granules; the ratio grows with granular diameter because the width
of the intergranular lanes is fixed.
Hence the ratio grows almost like the ratio of area to circumference, i.e.
nearly linearly with granule diameter.
Changing the width of the intergranular lanes while keeping the size
distribution fixed gives basically the same result.

Note that changing a single parameter can lead to subtle indirect changes in
the properties of the ensemble of granules and hence of the wavelet power
spectrum.
As an example, a change in the size distribution of granules also leads to a
change of their effective lifetime distribution.
Thus for a very small
average granule size more granules die early by reaching the
limits of the size distribution. For a star of constant size the
number of granules is reduced if the average granule size is
bigger. Therefore the curves in Fig.~\ref{modelvar3} would lie
even further apart for a fixed stellar surface area.
The dotted line would lie higher by a factor of approximately $\sqrt{3}$,
while the dashed line would move down by the same factor.
We also found that splitting granules on the stellar surface produce higher
levels of noise than dissolving granules.
This is partly due to the
fact that splitting granules are on average bigger, so that a
similar picture emerges as shown in Fig.~\ref{modelvar3} (with
some differences, since the shape of the lifetime and size
distributions change significantly with the ratio of splitting to
dissolving granules).

\section{Magnetic field contribution to irradiance variations}
\label{magnetic}


Here we employ the SATIRE-S \citep[Spectral And Total Irradiance
REconstruction for the Satellite
era;][]{fligge-et-al-2000a,krivova-et-al-2003a,solanki-et-al-2005a,%
krivova-et-al-2011a} model, which is based 
on the assumption that all irradiance changes on
time\-scales of days to years are entirely due to the evolution of
the magnetic flux on the solar surface and includes four components
of the solar photosphere: quiet Sun (solar surface free of magnetic
fields), umbra and penumbra of sunspots, as well as bright magnetic
features forming faculae and the network (described as a single
component).
Calculation of solar irradiance requires two input data sets.
The first one is the intensities of each atmospheric component
as a function of the wavelength and of the angle between the line of sight
and the normal to the solar surface (i.e. the center-to-limb variation).
These time-independent spectra are taken from \citet{unruh-et-al-99}.
The second data set is composed of the maps describing the distribution of
the magnetic features (umbrae, penumbrae, plage and faculae) on the solar
surface at a given time.
They are produced from magnetograms and continuum images (see Krivova et al.
2003 for details) recorded by MDI on board SOHO (Scherrer et al. 1995).
The changing area coverage and distribution of the different components
introduces the time variability of the brightness.

Usually, MDI does not record continuum images and full-disk magnetograms
with the necessary low noise level (5-min integration, corresponding to a
noise level of 9 G) at the cadence required for the present investigation.
However, we
found a 2.2-month period from 15th of March to 19th of May 1999 when MDI
magnetograms were recorded at a rate of at least one 5-min sequence every
half an hour practically without interruptions.
Unfortunately, the continuum images, required to
determine umbral and penumbral area, were recorded with a cadence of only 96
minutes over the same period of time.
Therefore, every continuum image is used for 3 points in time by rotating it
to the times of the corresponding magnetograms.
Since sunspots can evolve over
this interval, the obtained power at periods of a few hours is
probably somewhat underestimated.

SATIRE-S has a single free parameter, $B_{sat}$,
which enters in how the magnetogram signal in facular and network regions is
converted into brightness.
It is described in detail by, e.g.,
\citet{fligge-et-al-2000a,krivova-et-al-2003a,wenzler-et-al-2006a}.
Here we adopt the value of $B_{sat}$=280~G following
\citet{krivova-et-al-2003a}.

The time series produced in this manner is plotted in Fig.~\ref{recvir}
(lower black curve).
Compared with the irradiance fluctuations measured by VIRGO (top curve in
Fig.~\ref{recvir}) the irradiance variations induced by the magnetic field
follow the longer term variations very well, but are too smooth, i.e. they
obviously lack power at very short periods.
Some artifacts (small brightening spikes) are seen at selected times between
days 5 and 11.
They are due to artifacts in some of the images recorded on these days.
The power spectrum resulting from this reconstruction is represented by the
dotted line in Fig.~\ref{combirec_separate} (between about 1 and
250$\mu$Hz).

\section{Combined short and long time scale variability}
\label{combined}

   \begin{figure}
   \centering

\includegraphics[width=9cm]{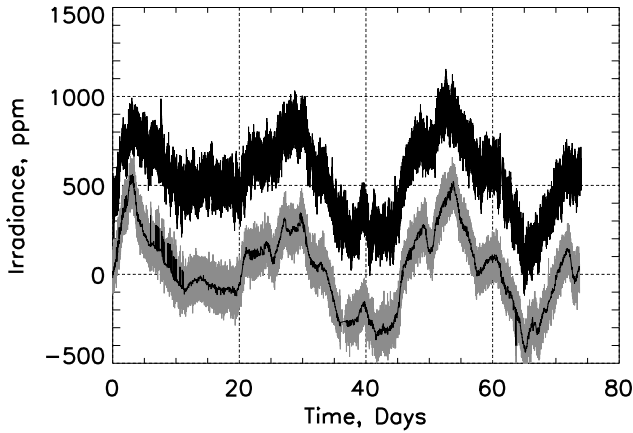}
      \caption{
Total solar irradiance observed by VIRGO between March and May 1999 (upper
black curve) and reconstructed using MDI data (lower black curve). Also
plotted (in gray) is the combination of magnetic reconstruction with
granulation model. The sampling rate of the VIRGO data is 1 minute, while
the magnetic reconstruction alone is sampled each 30 minutes. The magnetic
reconstruction and the combined reconstruction plus granulation model are
shifted down by 500~ppm to facilitate the visual comparison.
              }
         \label{recvir}
   \end{figure}

Now we have 2 parts of the solar irradiance~--- one is reconstructed
using only the magnetic activity of the Sun, the other is based
exclusively on the convective `noise' of the solar surface, i.e.
caused only by granulation.
In order to compare modelled variability and observed one, we combine both
magnetic and granulation parts.
The granulation model used here is the `standard' model with granule
parameters based on observed solar values (see Sect.~3.1 for a
description).
It corresponds to the solid curves in Figs.~\ref{modelvar1} and
\ref{modelvar3}.
Both models were run for the same length of time here, namely 2.2 months.
To combine magnetic reconstructions with granulation model results we
re-sample magnetic reconstructions from 30 minutes down to a 1 minute
sampling rate.
As seen in Fig.~\ref{recvir}, the combined model (gray curve) follows very
closely the observed one.

   \begin{figure}
   \centering

\includegraphics[width=9cm]{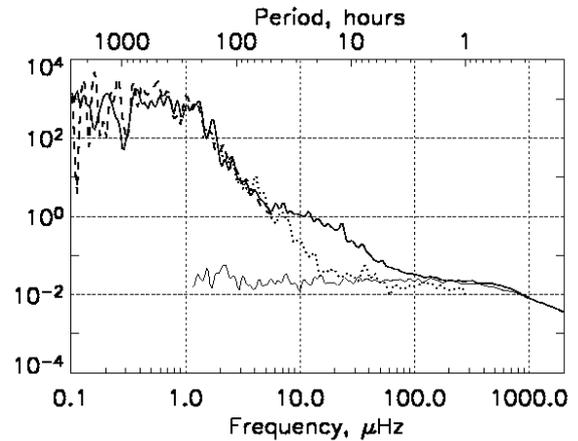}
      \caption{
 Wavelet power spectra
(in ppm$^2/60~\mu$Hz).
 Thick solid line: VIRGO data set (same as dotted line in Fig.~1).
 Thin solid line:
 power spectrum of the granulation model, dotted line: magnetic
 reconstruction with 30 minutes sampling, dashed line in the left part of
 the plot: daily sampled magnetic reconstruction.
              }
          \label{combirec_separate}
   \end{figure}

Power spectra of the modelled and observed irradiance allow a more detailed
comparison.
Figure~\ref{combirec_separate} shows the wavelet power spectra of the
observed data (thick solid line), of the irradiance produced by the
granulation model (high frequencies; thin solid line) and the magnetic field
model (frequencies $\approx 1-250\mu$Hz; dotted line).
Since half-hourly sampled magnetograms are available only for 2.2 months,
while the diagram covers 4 months (and VIRGO data allow even longer periods
to be considered), we have also employed the reconstruction carried out by
\citet{krivova-et-al-2003a} which is sampled at a cadence of 1 day and
represented by the dashed line in the left part of
Fig.~\ref{combirec_separate}.
The low frequency part, caused exclusively by the magnetic activity, is in
good agreement with the observed spectrum.
The high frequency part obtained using the solar surface granulation model
(now run for a larger number of granules) recreates real power rather well
at frequencies above 100~$\mu$Hz and gives a negligible contribution to the
combined irradiance at frequencies below 10~$\mu$Hz.
Figure~\ref{combirec_separate} confirms that the solar total irradiance
noise background is equally determined by convection (granulation) and
magnetism at periods of 10 hours (cf. Fig.~\ref{wavelet_ratio}).

\begin{figure}
   \centering
\includegraphics[width=9cm]{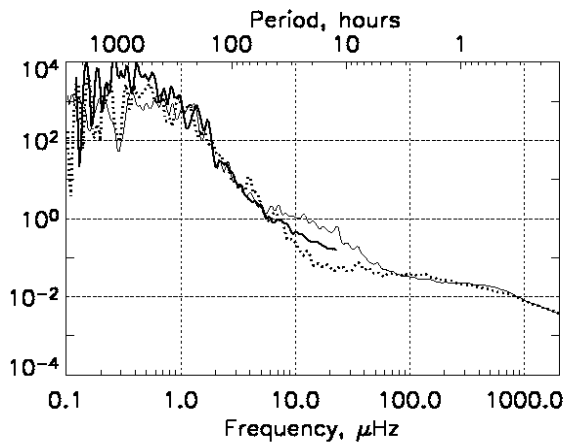}
      \caption{
      Wavelet power spectra (in ppm$^2/60~\mu$Hz).
      Thin solid line: VIRGO data set. Dotted line:
      power spectrum of the combined magnetic reconstruction and granulation
      model. Thick solid line: SORCE wavelet power spectrum.
              }
        \label{combirec}
   \end{figure}

In order to compare the observed and modelled power spectra in more detail,
we combine all 3 components of the model.
Figure~\ref{combirec} shows wavelet power
spectra of the observed VIRGO (thin solid line) and SORCE data
(thick solid line) and of the reconstructed irradiance (dotted)
produced by combining the various reconstructions plotted 
in Fig.~\ref{combirec_separate}. SORCE data
are for the year 2003 and sampled every 6 hours, while the
employed VIRGO data are sampled each minute and refer to 2002.
The comparison with the SORCE spectrum confirms that the VIRGO power
enhancement in the band from 7 to 30~$\mu$Hz is an artifact (C. Fr\"ohlich,
priv. comm.).
The different years to which the SORCE TIM and VIRGO data refer to should
not affect this result, since variability at 10--100~$\mu$Hz is hardly
affected by solar cycle phase (see Fig.~\ref{wavelet_ratio}).
Note, however, that the modelled variability lies below both VIRGO and SORCE
TIM measurements in the frequency range 8--70~$\mu$Hz.

\section{Conclusions}
\label{conclusions}

We have analyzed solar irradiance variations on time scales between minutes
and months.
Whereas on time scales of a day and longer, the main mechanism of variations
is the evolution of the solar magnetic field, on time scales shorter than
roughly half a day granular convection becomes dominant.
The crossover between magnetic and convective signatures coincides with the
frequency band of most interest for planetary transit observations.

If we neglect the 5-min band, then results of combined modelling of
irradiance variations due to granular convection and surface magnetism
suggest that we are able to reproduce solar irradiance variability using
only data of magnetic activity on the surface of the Sun and `noise'
produced by granulation, except for periods between 5 and 30 hours.
In this range of periods the solar variability is underestimated by the
model by up to a factor of 5 when comparing with VIRGO TSI and up to a
factor of 2 when comparing with SORCE TIM.
Due to the influence of electronic noise on the VIRGO data, we believe the
latter ratio is the more reliable one.
The discrepancy may have one of the following causes:
1. The lack of larger scales of convection, such as supergranulation (and
possibly mesogranulation) in the model.
2. Insufficient sensitivity of MDI magnetograms to weak fields (i.e. to
quiet Sun fields) and therefore also to their variability.
3. Insufficient spatial resolution of the magnetograms, so that flux with
mixed polarities at small spatial scales, as is typical of the quiet Sun,
can be missed \citep{krivova-solanki-2004a}.
4. The breakdown of one of the assumptions underlying the magnetic
field-based reconstructions.
For example, the simplifying assumption that all magnetic features at a
certain limb distance and a given magnetic flux have the same brightness is
expected to break down at some level.
5. The high frequency component of the evolution of sunspots and pores is
inaccurately represented by the interpolation between the continuum
images separated by 96 minutes.
The Nyquist frequency of this data set lies at 45~$\mu$Hz, making this
explanation less likely.
Note that the time-scale on which  the quiet Sun network flux is replaced by
ephemeral regions, 14~hours according to \citet{hagenaar-2001}, falls in the
critical time range and supports causes 2 and 3 above.

The present investigation also provides an evaluation of the diagnostic
potential of power spectra of radiative flux time series for determining the
properties of stellargranulation.
Granule contrasts, the number of granules,
granule lifetimes and diameters, as well as contrast and thickness of
intergranular lanes are important factors determining the amplitude and the
shape of the power spectrum at periods shorter than several hours.
In particular, power spectra, such as those that can be obtained from COROT
\citep{baglin-et-al-2002} and Kepler \citep{borucki-et-al-2003} data, allow
granule lifetimes to be determined in a relatively unique manner.
This is of some interest, since traditional techniques for diagnosing stellar
granulation, such as line bisectors, do not provide any information on
granule lifetimes.
Note that this diagnostic is not affected by any uncertainties in the model
below 100~$\mu$Hz.

Important next steps are to identify the source of the missing power around
periods of 10--20 hours and to extend such an analysis to other stars with
different effective temperatures and gravities, as well as with different
rotation rates and magnetic activity levels.
Magnetograms and continuum images obtained at high cadence by the HMI
instrument on SDO could play a prominent role in the former exercise.

\begin{acknowledgements}
We thank C.~Fr{\"o}hlich for providing data and clarifications
regarding the SoHO/VIRGO instrument.
We thank the SoHO/MDI team for providing access to magnetograms and continuum
images.
SoHO is a project of international cooperation between ESA and NASA.
Also we thank the SORCE team and in particular G.~Kopp for providing data
for comparison.
This work was supported by the
\emph{Deut\-sche For\-schungs\-ge\-mein\-schaft, DFG\/} project
number SO~711/1-1/2 and by  WCU grant (No. R31-10016) funded by the Korean
Ministry of Education, Science and Technology.
\end{acknowledgements}

\bibliographystyle{aa}

\end{document}